# Comprehending Quantum Theory from Quantum Fields

Mani Bhaumik[1]
Department of Physics and Astronomy,
University of California, Los Angeles, USA.90095

## Abstract

At the primary level of reality as described by quantum field theory, a fundamental particle like an electron represents a stable, discrete, propagating excited state of its underlying quantum field. QFT also tells us that the lowest vacuum state as well as the excited states of such a field is always very active with spontaneous, unpredictable quantum fluctuations. Also an underlying quantum field is known to be indestructible and immutable possessing the same value in each element of spacetime comprising the universe. These characteristics of the primary quantum fields together with the fact that the quantum fluctuations can be cogently substantiated to be quantum coherent throughout the universe provide a possible ontology of the quantum theory. In this picture, the wave function of a quantum particle represents the reality of the inherent quantum fluctuations at the core of the universe and endows the particle its counter intuitive quantum behavior.

## 1. Introduction

Despite the indisputable success of the quantum theory for nearly a century, the existence of an objective quantum reality has been a subject of contentious debates, particularly as regards to the reality of the wave function itself. Almost since the inception of the theory, the question as to whether the wave function represents an element of reality or merely gives a subjective knowledge of the system has been in dispute. Early on, the subjective knowledge paradigm dominated among the physicists, although Einstein stubbornly refused to accede. Instead he championed the famous EPR paradox [1]. David Bohm also eschewed the subjective interpretation. As an alternative, he tenaciously pursued the concept of an observer-independent element of objective reality with his 'quantum mechanical potential' and some unspecified 'hidden variables'. [2]. But his impromptu invocation of the quantum mechanical potential, subsequently called simply quantum potential, made his theory look somewhat artificial, relegating it to relative obscurity until years later when John Bell [3] gave it substantial support. Bell's pioneering work led to affirmation of non-locality implicit in Bohm's quantum potential proposition, but also to the demise of any local hidden variable explanation for quantum behaviors. Most importantly, Bell's breakthrough resulted in the astounding discovery of quantum entanglement as well as the actuality of superposition of quantum states.

---

[1] e-mail: bhaumik@physics.ucla.edu

In their latest paper, Colbeck and Renner [4] advanced some logical arguments in support of the element of reality of the wave function. However, neither Bohm nor the authors of the recent paper presented any foundation for the existence of such a reality. Contemporary experimental observations supported by the quantum field theory demonstrate the immutability and a universal value everywhere of an underlying quantum field at the primary level of reality as well as some definitive confirmation of the existence of inherent fluctuations of the quantum field. These profound fundamentals of reality together with the consequences of the observed phenomenon of quantum entanglement appear to support the objective element of reality of the wave function.

## 2. Nature of Primary reality portrayed by quantum field theory

The quantum field theory has uncovered a fundamental nature of reality, which is radically different from our daily perception. Our customary ambient world is very palpable and physical. But QFT asserts this is not the primary reality. The fundamental particles involved at the basis of our daily physical reality are only secondary. They are excitations of their respective underlying quantum fields possessing propagating states of discrete energies, and it is these which constitute the primary reality. For example, an electron is the excitation of the abstract underlying electron quantum field. This holds true for all the fundamental particles, be a boson or a fermion. Inherent quantum fluctuations are also a distinct characteristic of a quantum field. Thus, QFT substantiates the profoundly counter intuitive departure from our normal perception of reality to reveal that the foundation of our tangible physical world is something totally abstract, comprising of discrete quantum fields that possess intrinsic, spontaneous, unpredictable quantum fluctuations.

It would be cogent to ask how we know that these quantum fluctuations really exist. Since a quantum system has to be irrevocably disturbed to observe it, we normally look for their evidence *indirectly* through their effects such as the Casimir effect, Lamb Shift and a host of other phenomena. Perhaps the most spectacular example of their existence is revealed in the extensive studies of the electron gyromagnetic ratio, where the experimentally observed value agrees with the calculated value based on QFT to an astonishing accuracy of one part in a trillion. Moreover in recent times, a graphic demonstration of their existence seems to have been provided by nature itself in the way of minute temperature inhomogeneities in the Cosmic Microwave Background Radiation (CMBR) Fig 1. It is now generally accepted that these inhomogeneities in the CMBR most likely owe their origin to quantum fluctuations of an inflaton field, manifesting as wrinkles in spacetime and blown up from their microscopic existence to macroscopic dimensions by a sudden explosive expansion of space in the very early universe, known as inflation. These exceptional depictions are considered to provide compelling evidence of the real existence of the quantum fields and their innate fluctuations.

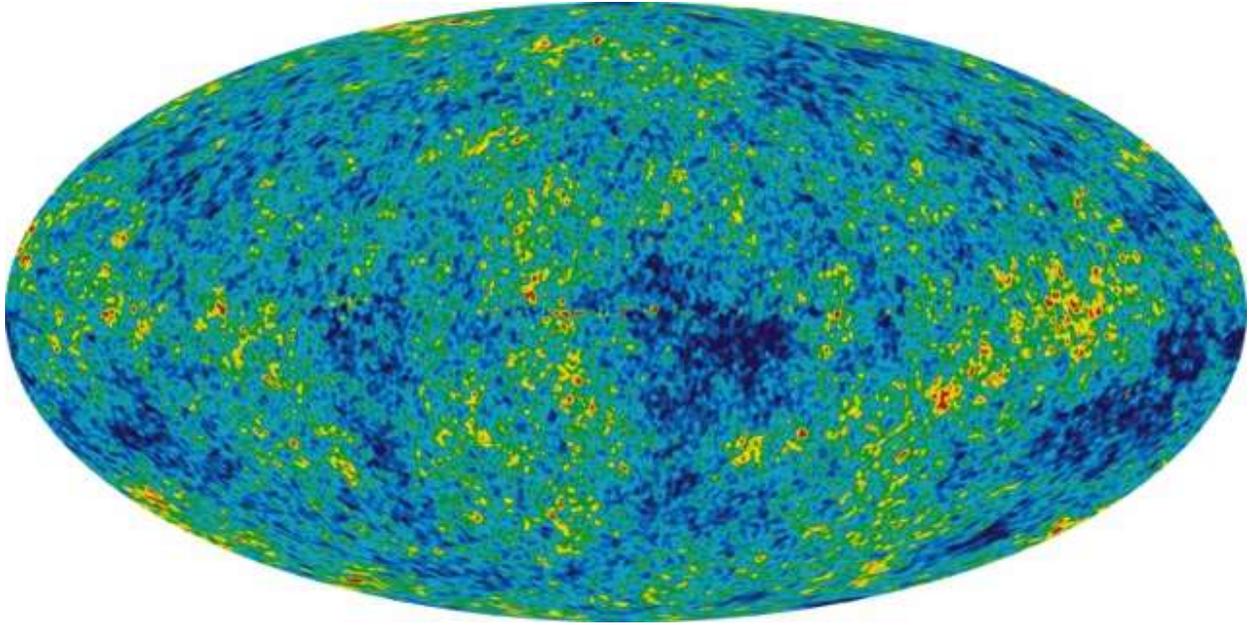

Figure 1. The picture made available by WMAP team shows the minute temperature inhomogeneites, about 1 part in 100,000, in the Cosmic Background Radiation observed by the WMAP satellite. These inhomogeneities most likely owe their origin to the fluctuations of the quantum field and are also believed to provide seeds for forming galaxies, galactic clusters and super clusters.

By far, the most phenomenal step forward made by QFT is the stunning prediction that the primary ingredient of *everything* in this universe is present in *each element of spacetime* of this immensely vast universe [5]. These ingredients are the underlying quantum fields. We also realize that the quantum fields are alive with quantum activity. These activities have the unique property of being completely spontaneous and utterly unpredictable as to exactly when a particular event will occur cannot be predicted [5]. But even to use a word like 'event' renders this activity in slow motion. In actuality, some of the fluctuations occur at mind-boggling speeds with a typical time period of $10^{-24}$ second or less. In spite of these infinitely dynamic, wild fluctuations, the quantum fields have remained immutable possessing the same magnitude, as evinced by their Lorentz invariance, essentially since the beginning and throughout the entire visible universe containing regions, which are too far apart to have any communication even with the speed of light. This is persuasively substantiated by the experimental observation that a fundamental particle such as an electron has exactly the same properties, be its mass, charge or spin, irrespective of when or where the electron has been created, whether in the early universe, through astrophysical processes over the eons or in a laboratory today anywhere in the world.

As recently explained by Narnhofer and Thirring [6], in quantum field theory almost everything is entangled. This essentially paints a broad picture at least at fundamental dimensions as well as for atomic scales. As a consequence, the fluctuations of the fields in each element of spacetime are expected to be quantum coherent. This is also corroborated by Frank Wilczek [7]. Then all the fluctuations of the fields would be coherent throughout the universe by mesoscopic quantum entanglement [8]. QFT substantiates existence of abundant interactions to produce various degrees of such entanglement. Therefore, the observed picture of primary reality supported by QFT comprises of the underlying immutable quantum fields and their quantum coherent fluctuations in a universal scale. *A fundamental particle arising out of its quantum field is always inescapably subject to this reality and is thus describable by an associated wave function that corresponds to actuality.*

## 3. The double slit Experiment

Let us take as an example the legendary double slit experiment, which Richard Feynman called both "the heart of quantum mechanics" and it's most enduring mystery. Specifically, we examine the results presented in Fig. 2, which was obtained by shooting electrons one at a time in succession to a double slit and observed on a screen. One of the results of such an experiment demonstrates that the final pattern 2e turns out to be the same whether the electrons are shot one at a time or all at the same time. The fact that the final pattern is independent of time spacing is classically incomprehensible. It suggests that there is an indestructible underlying entity, which is responsible for this behavior. This is exactly what the quantum field theory says─what underlies is the immutable electron quantum field which fills all of spacetime.

We can explore the same underlying quantum field to uncover the reason for the other classically incomprehensible aspect of this double slit experiment. As shown in the Fig. 2, we conspicuously find that each particle makes an unpredictable choice between alternate possibilities. Again, recalling our earlier discussions that according to QFT and supported by experimental observations, a fundamental particle represents a stable, discrete propagating excited state of its underlying quantum field. QFT tells us further that the spontaneous and unpredictable quantum activity is also present *in the excited state*. Mathematical analysis presented here shows this quantum activity could endow the particle its observed property of making an apparently unpredictable choice.

For the sake of simplicity we will consider the motion of a single nonrelativistic electron without spin. To a good approximation, in QFT the wave function of quantum fluctuations can be represented by a linear superposition of harmonic oscillator wave functions. The wave function $\psi$ of quantum fluctuations can be written as

$$\psi = \sum_i c_i \psi_i$$

The function $\psi$ can be expressed in polar coordinates as

$$\psi = R \exp(iS/\hbar),$$

where $R$ is the amplitude and $S/\hbar$ is its phase, both $R$ and $S$ being real.

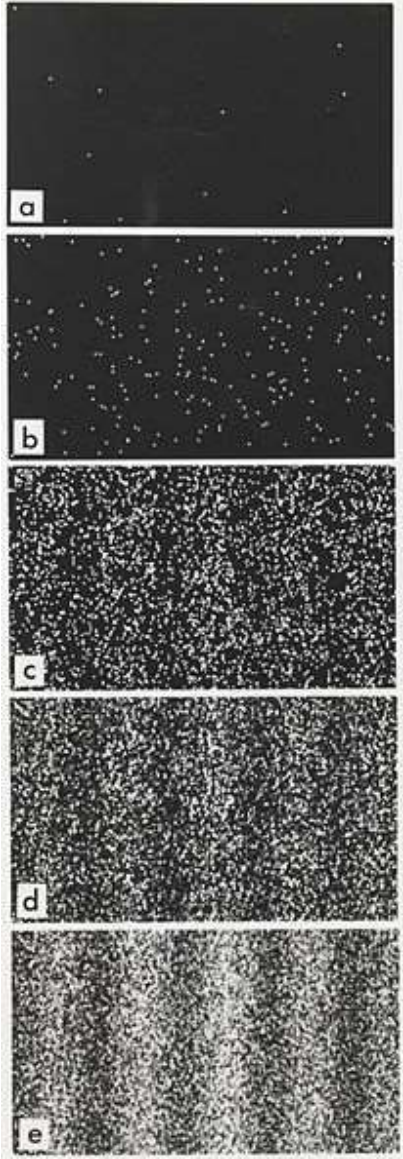

Figure 2. The results of a double-slit experiment performed by Dr. Tonomura by shooting one electron at a time in succession to the slits. Images **a** through **e** depict collection of gradually increasing number of electrons on the screen. Numbers of electrons are 10 (a), 200 (b), 6000 (c), 40000 (d), 140000 (e). [9]

Since $\psi$ should be a solution, following Bohm [2,10,11], we insert $\psi$ in the Schrödinger equation for a single particle:

$$i\hbar \frac{\partial \psi}{\partial t} = -\frac{\hbar^2}{2m} \nabla^2 \psi + V\psi$$

where *V* is the classical potential.

The equation then factors in to two equations, one imaginary and the other real.

The real part of the Schrödinger equation in polar form yields

$$\frac{\partial S}{\partial t} + \frac{(\nabla S)^2}{2m} + V - \frac{\hbar^2}{2m}\frac{\nabla^2 R}{R} = 0$$

It essentially differs from the classical Hamilton–Jacobi equation only by the term

$$Q = -\frac{\hbar^2}{2m}\frac{\nabla^2 R}{R}$$

where $Q$ is called the *quantum potential* by Bohm [2]. Bohm and Hiley [10, 11] have carried out extensive investigations on it over the years and have successfully extended the quantum potential approach to many body systems. They emphasize several key aspects of their analysis involving the quantum potential [10, 11]:

1. The electron acts like a particle with a well-defined position which varies continuously.
2. The wave function $\psi$ of an individual electron is regarded as a mathematical representation of an objectively real field.
3. The particle is never separated from this new type of quantum field that fundamentally affects it. The field $\psi$ satisfies the Schrödinger equation so that it too changes continuously.
4. Given that the particle is always accompanied by its quantum field, the combined system of the particle and the field is causally determined.
5. The quantum field represents an internal energy without a source and determines the magnitude of the physically significant quantum potential.
6. The quantum potential Q provides the quantum force $-\nabla(Q)$ for the equation of motion of the particle:

$$m\frac{dv}{dt} = -\nabla(V) - \nabla(Q).$$

7. The quantum potential does not fall off with distance, implying nonlocality.

All of these aspects can be cogently supported by our earlier description of the Lorentz invariant, immutable underlying *primary* quantum field that fills all space and time with the same magnitude as well as the quantum coherence of its intrinsic fluctuations, thereby providing the crucial element of reality for the quantum potential that has been missing from Bohm and Hiley's treatment.

In the double slit experiment, the electron approaches the slits along with its inseparable quantum wave. While the electron would presumably go through one slit, the accompanying quantum wave, by its interaction with the constituents of the slits, can go through both slits. However, transition through both slits affects the wave to produce a compound quantum potential that changes the trajectory of the electron. Philippides, et al [12] calculated the trajectories of the electrons as adjusted by the compound quantum potential. They substantiated that the modified trajectories support the observation that each particle makes an unpredictable choice between alternative possibilities, consistent with an interference pattern. In other words, the inherent fluctuations of the underlying quantum field render the electron to choose an arbitrary outcome from alternate possibilities.

When an electron lands on the screen, the wave function diffuses by decoherence. Such decoherence is facilitated by interaction of the electron's wave function with the atoms of the detector having irreversible random thermal motions. When the totality of the trajectories of many electrons terminates on the screen, the resulting configuration corresponds [12] to an interference pattern similar to that shown in Fig. 2e.

Extending Bohm and Hiley's quantum potential approach, we therefore present here a persuasive paradigm indicating the wave function of a quantum particle represents the reality of the innate quantum fluctuations at the core of the universe and bestows the particle its uncanny quantum behavior, thereby providing a credible ontology for the quantum theory. In contrast to Bohm's original presentation, we have eliminated the need for the experimentally refuted local "hidden variables" as well as the need for invocation of a quantum potential without foundation.

## 4. Acknowledgement

The author wishes to thank Professor Walter Thirring, Professor Frank Wilczek, and Professor Zvi Bern for helpful discussions.